\documentclass[preprint,proceedings]{rmaa}

\suppressfulladdresses 

\usepackage{paralist}

\usepackage{psfrag,color}

\newcommand{\nlte} {{\sc nlte}}
\newcommand{\lte} {{\sc lte}}
\newcommand{\sed} {{\sc sed}}
\newcommand{\fastwind} {{\sc fastwind}}

\newcommand{\tlusty} {{\sc tlusty}}

\newcommand{\cloudy} {{\sc cloudy}}
\newcommand{\vsini} {$v${\thinspace}sin{\thinspace}$i$}
\newcommand{\kms} {km\,s$^{-1}$}
  
\SetYear{2007}
\SetConfTitle{Massive Stars: Fund. Param. and Circumst. Inter.}

\title{Massive stars and their surrounding nebulae: A combined approach} 

\author{
  S. Sim\'on-D\'iaz,\altaffilmark{1} 
  G. Stasi\'nska,\altaffilmark{1}
  J. Garc\'ia-Rojas,\altaffilmark{2}
  C. Morisset,\altaffilmark{3}
  A. R. L\'opez-S\'anchez,\altaffilmark{2}
  and C. Esteban\altaffilmark{2}}

\altaffiltext{1}{LUTh, Observatoire de Meudon, France.}

\altaffiltext{2}{Instituto de Astrof\'\i sica de Canarias, Spain.}

\altaffiltext{3}{Instituto de Astronom\'ia, Universidad Nacional 
Aut\'onoma de M\'exico, M\'exico.}

\shortauthor{Sim\'on-D\'iaz et al.}
\shorttitle{Massive stars and H\,II regions: a combined approach}

\listofauthors{S. Sim\'on-D\'iaz, G. Stasi\'nska, J. Garc\'ia-Rojas,
  C. Morisset, A. R. L\'opez-S\'anchez, \& C. Esteban}
\indexauthor{Sim\'on-D\'iaz, S.}
\indexauthor{Stasi\'nska, G.}
\indexauthor{Garc\'ia-Rojas, J.}
\indexauthor{Morisset, C.}
\indexauthor{L\'opez-S\'anchez, A. R.}
\indexauthor{Esteban, C.}

\abstract{We present the first results of a project aimed at the combined study 
of massive stars and their surrounding nebulae by means of a detailed study of 
Galactic H\,{\sc ii} regions ionized by only one massive star. With this, we
intend to check the validity of the new generation of massive star model 
atmosphere codes in terms of ionizing flux distribution. We take into account
the effect of the nebular density distribution in our analyses. Various 
types of stellar and nebular observations have been collected for this purpose.}

\resumen{Presentamos los primeros resultados de un proyecto dedicado al estudio 
combinado de estrellas masivas y sus regiones H\,{\sc ii} asociadas en la Galaxia. 
Mediante el estudio de regiones ionizadas por una \'unica estrella pretendemos 
comprobar la validez de las predicciones de la nueva generaci\'on de modelos de 
atm\'osfera de estrellas masivas en t\'erminos de distribuci\'on de flujo 
ionizante. En nuestros an\'alisis consideramos el efecto de las posibles 
distribuciones de densidad del material ionizado. Para ello hacemos uso de 
varios tipos de observaciones, tanto estelares como de la regi\'on
H\,{\sc ii}.}

\addkeyword{Stars: early-type}
\addkeyword{Stars: individual: HD\,37061}
\addkeyword{H~II regions}
\addkeyword{ISM: ionization models}
\addkeyword{ISM: individual objects: M\,43}

\begin{document}
\maketitle

\section{Introduction}
\label{sec:intro}

The intense far ultraviolet radiation emitted by early OB-type stars ionizes 
the interstellar medium, generating the so-called \ion{H}{ii} regions. These 
ionized regions can be used to derive properties of the associated stellar 
population (e.g. star forming rates, age). However, since the properties of 
\ion{H}{ii} regions crucially depend on the spectral energy distribution 
(\sed) of the massive star population, and this part of the stellar flux is 
generally unaccesible to direct observations, the predictions resulting from 
massive star atmosphere codes are a crucial ingredient.

The new generation of \nlte, line blanketed model atmosphere codes
(Hubeny \& Lanz, 1995, Hillier \& Miller 1998, Pauldrach et al. 2001,
Puls et al. 2005), which include a more realistic description of
the physical processes characterizing the stellar atmosphere, produce 
quite different ionizing \sed s than the previous plane-parallel, 
\nlte/\lte, hydrostatic models (Mihalas \& Auer 1970, Kurucz 1992, Kunze 1994). 
Some notes on this, and on the consequences on the ionization structure
of \ion{H}{ii} regions, can be found in Gabler et al. (1989), Rubin et al.
(1995), Najarro et al. (1996), Sellmaier et al. (1996) and Stasi\'nska 
\& Schaerer (1997).
Although new predictions seem to go in the right direction (viz. Giveon et 
al. 2002, Morisset et al. 2004), non-negligible differences can be found 
between the different stellar atmosphere codes (viz. Martins et al. 2005, 
Puls et al. 2005).\\
\newline
We are performing a study of Galactic H\,{\sc ii} regions ionized by 
only one massive star to check the validity of the ionizing \sed s
predicted by the new generation of massive star model atmosphere codes.
In forthcoming papers we will present the complete combined study 
of the various nebulae and their ionizing stars; here we show some 
preliminary results from our study of M\,43.

\section{Observational dataset}
\label{sec:refs}
Stellar and nebular observations were carried out with 
the Wide Field Camera ({\sc wfc}; narrow-band imaging in H$\alpha$, 
H$\beta$, [\ion{O}{iii}] and [\ion{S}{ii}]) and the Intermediate 
Resolution Spectrograph ({\sc ids}; spectroscopy of the 
ionizing star in the range 4000\,-\,5000\,\AA\,+\,the H$\alpha$ region, 
and long slit nebular spectroscopy in the range 3600\,-\,9700\,\AA); 
both instruments attached to the Isaac Newton Telescope ({\sc int}).
The long slit nebular observations were divided into smaller apertures 
along the nebular radii.

\begin{figure*}[!t]
\begin{minipage}{9.5cm}
\includegraphics[width=6.8 cm,angle=90]{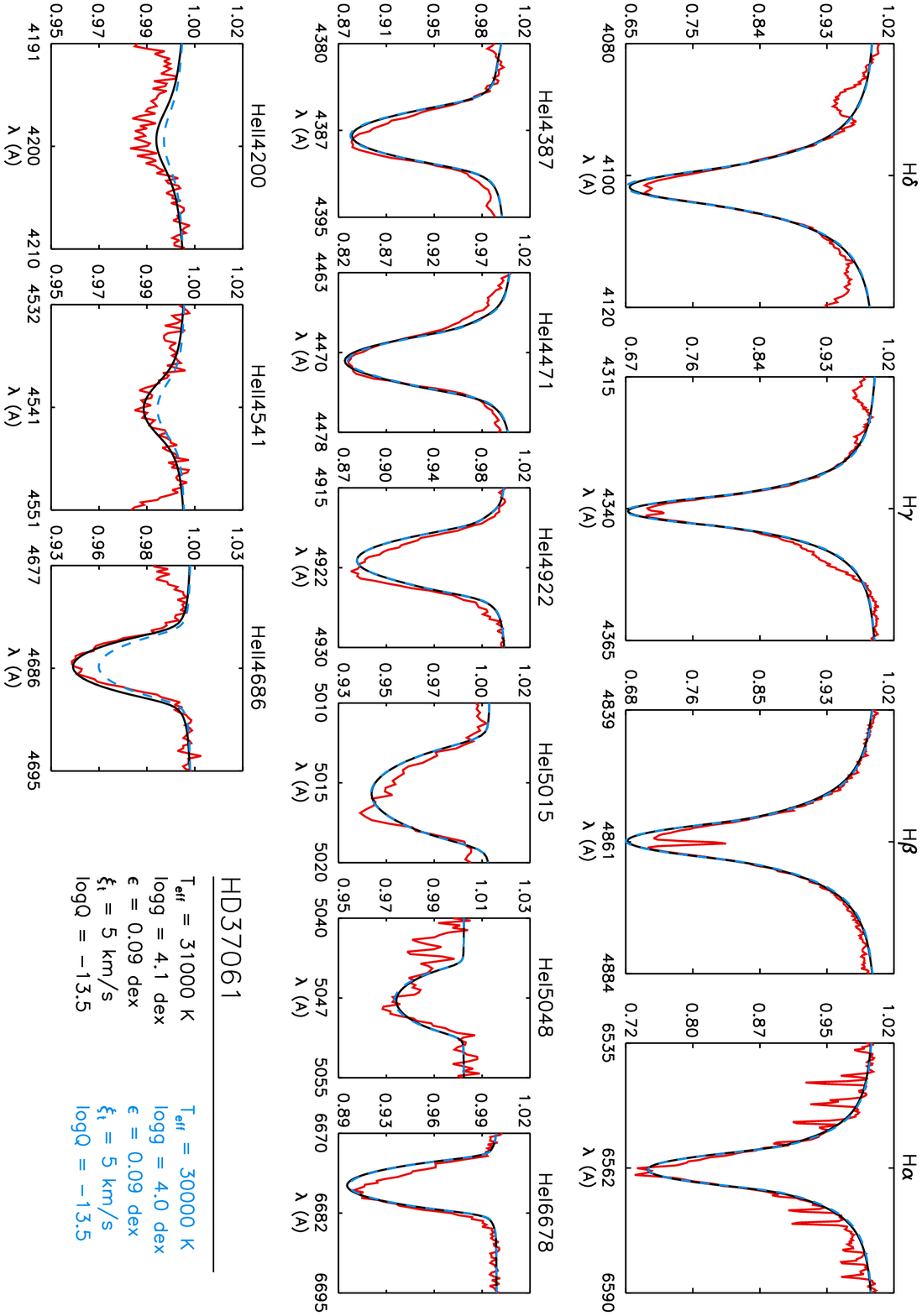}
\caption{\footnotesize
Fitting of \fastwind\ synthetic profiles (broadened to 
\vsini\,=\,210 \kms) to the optical H and He lines of HD\,37061. 
Two sets of stellar parameters are shown to illustrate the accuracy
of the stellar parameters determination.}
\label{fig1}
\end{minipage}
\   \
\hfill \begin{minipage}{6.5cm}
\centering
\includegraphics[width=5.0 cm,angle=90]{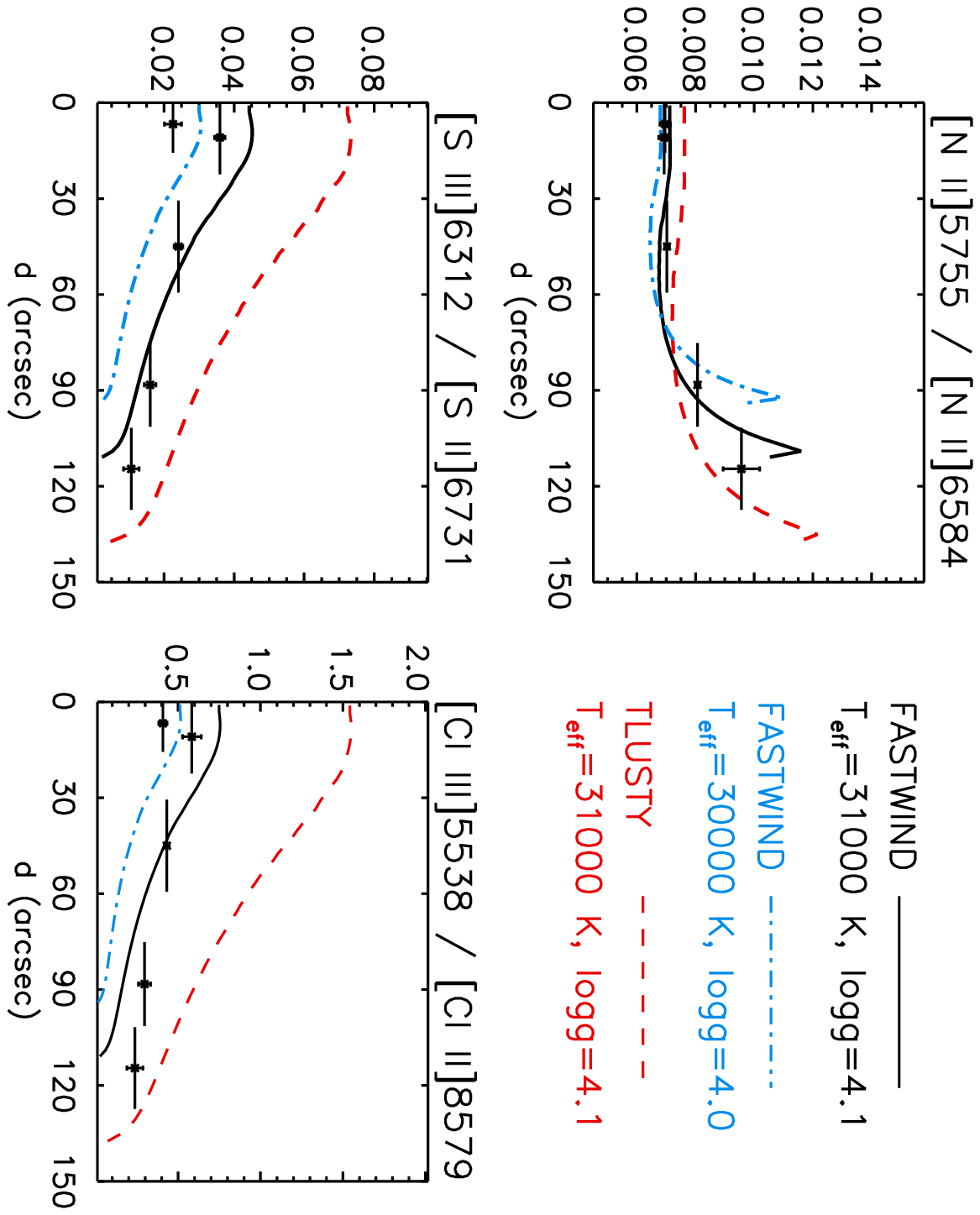}
\caption{\footnotesize
Spatial variation of various nebular line ratios resulting from 
three \cloudy\ constant density spherical models. Each model 
uses a different predicted \sed. \fastwind\ models are the ones 
used in Figure \ref{fig1} (same color code); a \tlusty\ model with the 
same stellar parameters as the \fastwind model which best fits the optical
HHe lines (Figure \ref{fig1}) has also been considered for comparison. Nebular 
observations from long slit spectra are also included.}
\label{fig2}
\end{minipage}
\end{figure*}

\section{Stellar and nebular analyses}
\label{sec:analyses}

M\,43 is an apparently spherical H\,{\sc ii} region, ioni\-zed by 
HD\,37061 (classified as B1\,V, though our spectrum clearly 
shows the \ion{He}{ii}\,4541 line, indicating that the star is rather 
B0\,V\,-\,B0.5\,V). The stellar 
parameters of HD\,37061 were obtained by visual fitting of 
\fastwind\ (Puls et al. 2005) synthetic profiles to the optical H and He lines  
(see Figure \ref{fig1}). 
Once the \sed\ resulting from the 
\fastwind\ model was re-scaled to fit $M_{\rm v}$\,=\,3.5 (from 
the stellar photometry and considering a distance of 450 pc), a
value of log\,$Q$(H$^0$)\,=\,47.2 could be derived. This later value is in good 
agreement with the nebular H\,$_{\alpha}$ luminosity calculated
from the {\sc wfc} H\,$_{\alpha}$ image, 
indicating that the nebula is ionization bounded.
From the nebular [\ion{S}{ii}]\,6731/6716 line ratio we inferred 
$N_{\rm e}$\,=\,550 cm$^{-3}$. In a first approach, we
have used the nebular abundances derived by Rodr\'iguez (1999)

\section{Photoionization models}
\label{sec:photomodels}

Figure \ref{fig2} illustrates three of the diagrams we use to 
compare the predictions of photoionization models (\cloudy, Ferland 
et al. 1998) with the observed nebular constraints. The first one is
an indicator of the nebular temperature, while the other two illustrate
the nebular ionization structure. 
These plots compare three constant density spherical 
\cloudy\ models: two of them take the \sed s from \fastwind\ models 
with different stellar parameters, while the third one was calculated 
using the \sed\ from a \tlusty\ (Hubeny \& Lanz, 1995) model with the 
same parameters as the \fastwind\ model which best fits the optical 
HHe lines. 
The differences between the photoionization models are important. 
Note, however, that geometries other than spherical and non-constant 
density distributions can affect the result (Morisset et al. 2005).
This will be also taken into account when inferring which model
better fit the nebular constraints. 
In forthcoming papers we will present the complete analysis in detail.

\end{document}